\documentclass[5p]{elsarticle}
\makeatletter
\def\ps@pprintTitle{%
 \let\@oddhead\@empty
 \let\@evenhead\@empty
 \def\@oddfoot{}%
 \let\@evenfoot\@oddfoot}
\makeatother
\usepackage{lineno,hyperref}
\usepackage{amsmath} 
\usepackage[font={scriptsize}]{subcaption} 
\usepackage{bm} 
\modulolinenumbers[1]
\graphicspath{{./figures/}} 
\journal{Nuclear Instruments and Methods in Physics Research A}

\begin{document}

\begin{frontmatter}

\title{Multiplication and Presence of Shielding Material from Time-Correlated Pulse-Height Measurements of Subcritical Plutonium Assemblies}

\author[UM_address]{Mateusz Monterial\corref{mycorrespondingauthor}}
\cortext[mycorrespondingauthor]{Corresponding author}
\ead{mateuszm@umich.edu}

\author[Sandia_address]{Peter Marleau}

\author[UM_address]{Marc Paff}

\author[UM_address]{Shaun Clarke}

\author[UM_address]{Sara Pozzi}

\address[UM_address]{Department of Nuclear Engineering and Radiological Sciences, University of Michigan, Ann Arbor, MI 48109, USA}
\address[Sandia_address]{Radiation and Nuclear Detection Systems Division, Sandia National Laboratories, Livermore, CA 94551, USA}

\begin{abstract}
We present the results from the first measurements of the Time-Correlated Pulse-Height (TCPH) distributions from 4.5 kg sphere of $\alpha$-phase weapons-grade plutonium metal in five configurations: bare, reflected by 1.27 cm and 2.54 cm of tungsten, and 2.54 cm and 7.62 cm of polyethylene. A new method for characterizing source multiplication and shielding configuration is also demonstrated. The method relies on solving for the underlying fission chain timing distribution that drives the spreading of the measured TCPH distribution. We found that a gamma distribution fits the fission chain timing distribution well and that the fit parameters correlate with both multiplication (rate parameter) and shielding material types (shape parameter). The source-to-detector distance was another free parameter that we were able to optimize, and proved to be the most well constrained parameter. MCNPX-PoliMi simulations were used to complement the measurements and help illustrate trends in these parameters and their relation to multiplication and the amount and type of material coupled to the subcritical assembly. 
\end{abstract}

\begin{keyword}
Fission chain \sep Liquid scintillator\sep Subcritical assembly\sep Non-linear Inverse Problem
\end{keyword}

\end{frontmatter}


\section{Introduction}

\subsection{Motivation and the new approach}

The ability to distinguish between multiplying and non-multiplying sources is of great interest to the safeguards and non-proliferation community. Neutron coincidence counting is one of the primary techniques in characterizing special nuclear material (SNM) properties such as presence of oxides, fissile mass, and by extension multiplication \cite{Ensslin1998}. We have focused on cross-correlation measurements with liquid organic scintillators as a possible method for characterizing SNM \cite{Enqvist2008, Miller2011}. The end-goal is a system that is smaller and more portable than traditional  thermal neutron capture based multiplicity counters. In addition, the method we propose would make accurate knowledge of system efficiency superfluous for the purposes of characterizing SNM. In this paper, we present a new technique for using inter-event timing data that could be collected by such a system. 

The Time-Correlated Pulse-Height (TCPH) distribution from measurements of the Beryllium Reflected Plutonium (BeRP) \cite{Mattingly2009} ball under various reflector configurations is presented in this work. We also present a method for inversely solving parameters that describe the fission chain timing distribution. Our underlying assumption is that for a multiplying source, such as the BeRP ball, the TCPH distribution is smeared by the distribution of times between fission events in a chain.  This fission chain timing distribution is shown to be sensitive to reflector composition, which complicates an absolute measurement of multiplication. 

A physical model of the fission chain distribution is beyond the scope of this paper. Instead, we used a gamma distribution as a proxy for the time smearing caused by the development of fission chains.  This approach has allowed us to draw empirical conclusions between the gamma distribution fit parameters and physical variations in fissile material assemblies, such as multiplication and coupling with a neutron moderating or reflecting material. 

We define the difference between moderating and reflecting materials by the relative energy loss of elastically scattering neutrons. Neutrons can scatter off of a high-Z reflector material at large angles without significant energy loss, and therefore quickly, on the order of nanoseconds, return to the multiplying medium. By contrast, in low-Z moderators scattering neutrons slow down significantly and may return to multiplying medium several microseconds later. 

The utility of the TCPH distribution has been extensively studied with non-multiplying $^{252}$Cf \cite{Miller2012}, low-multiplying mixed oxide powder and plutonium-gallium disks \cite{Miller2013}, and highly enriched uranium \cite{Marleau2013} measurements and simulations \cite{Paff2014}. BeRP ball experiments measuring the TCPH as a signature have only thus far been simulated, and these results demonstrated a correlation between multiplication and spreading in the TCPH distribution \cite{Monterial2013}. These earlier efforts to characterize multiplication relied on counting the correlated events that fell above the maximum theoretical arrival time of neutrons as calculated by their deposited energy. In this work, we advance beyond this methodology by reconstructing the time distribution between fission events and using this new signature for characterizing multiple aspects of a subcritical assembly.

\subsection{TCPH distribution and relation to fission chain timing}

Cross-correlation measurements combined with the pulse shape discrimination (PSD) methods allow for the classification of neutron-neutron, gamma-neutron, neutron-gamma, or gamma-gamma events \cite{Kno2000, Adams1978}. The TCPH distribution is a bivariate histogram of the time difference between correlated neutrons and gamma-rays, and deposited neutron energy measured by the light output in the scintillator. In a non-multiplying source, such as $^{252}$Cf, most correlated neutrons and gamma-rays detected within a short time window (10s of nanoseconds) by our fast system are generated from the same fission event. Therefore, the arrival time of a neutron, which depends on neutron incident energy, sets an upper limit on the light output achievable for that event. 

\begin{figure}[h!] 
 \centering
 \begin{subfigure}[h]{85mm} 
		\centering
		\includegraphics[width=\textwidth]{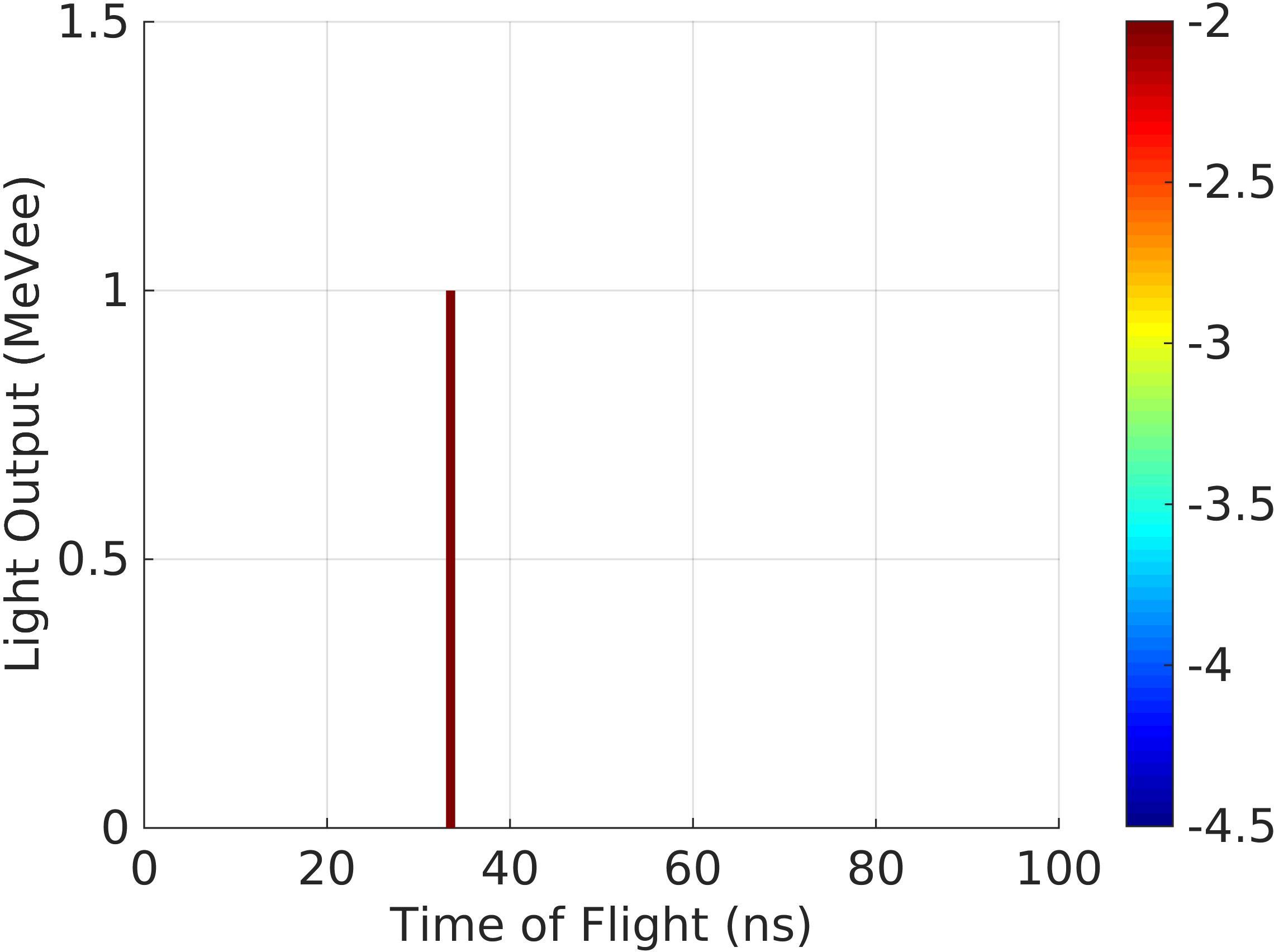}
 		\caption{Non-multiplying}
  \end{subfigure}%
     
        ~ 
  \begin{subfigure}[h]{85mm} 
		\centering
	  \includegraphics[width=\textwidth]{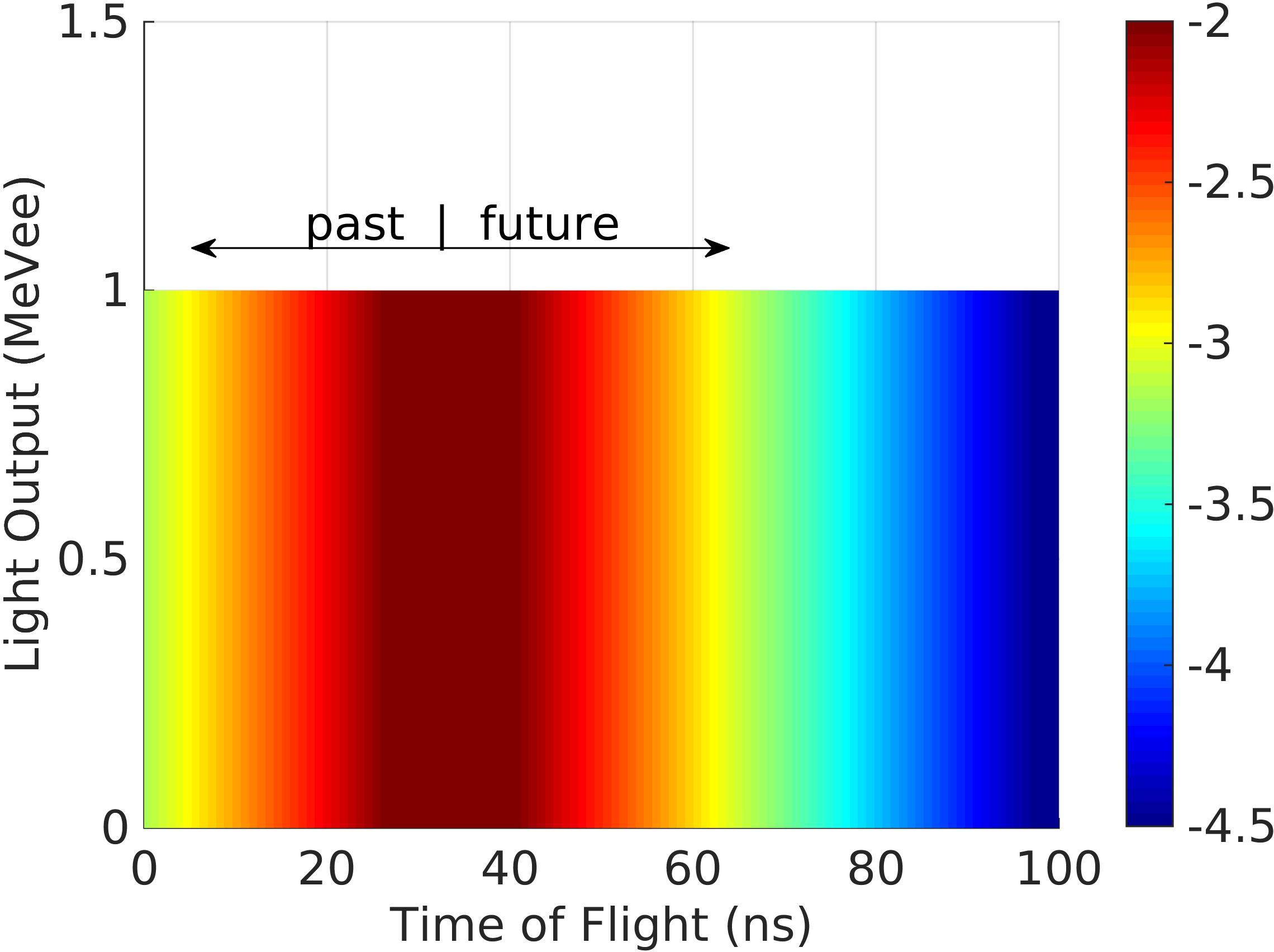}
   	\caption{Multiplying}
  \end{subfigure}%
  \caption{Calculated TCPH distributions for (a) non-multiplying and (b) multiplying sources assuming a mono-energetic 1 MeV neutron source at a distance of 50 cm with linear detector response and equal probability of scatter in all angles. Each distribution is normalized to unity and displayed on a logarithmic scale.}
  \label{fig:tcph_example}
\end{figure}

Figure \ref{fig:tcph_example}(a) shows an example of a TCPH distribution from a hypothetical non-multiplying mono-energetic source of 1 MeV neutrons and correlated gamma-rays. In this example, we assume the detector response is linear and all scattering angles are equally likely. In this case, all neutrons arrive 35 nanoseconds after the correlated gamma-ray due to the travel time of a 1 MeV neutron across a source-detector distance of 50 cm. In reality the arrival time would be slightly longer because of the extra collision distance within the detector. For the purposes of this example we assume that all interactions take place at the face of this hypothetical detector. 

For multiplying sources, the presence of fission chains makes it possible for neutrons and gamma-rays to be correlated in different events along the fission chain. The time of arrival is therefore either shorter or longer depending on the order of generation along the fission chain of the correlated gamma-neutron pair. In a multiplying assembly of fissile material this effect manifests itself as a spreading or smearing of the TCPH distribution. An example of the effect of this time smearing caused by source multiplication is shown in Figure \ref{fig:tcph_example}(b). The spreading in the negative and positive time direction is caused by the possibility of correlating gamma-rays with neutrons from past and future generations of fission events, respectively.

Gamma-gamma and neutron-neutron pairs could also be used to create TCPH-like distributions. However, these types of correlated events suffer from significantly more detector cross-talk as the same particle scatters twice in the detection system. Additionally, in neutron-neutron pairs, both particles have a time-of-flight that depends on their respective energies and thus the time between neutrons is spread even for pairs originating from the same fission. This decreases the sensitivity of the TCPH approach to time spread due to the presence of fission chains.

By contrast, correlated gamma-rays from the same fission should arrive at the detector simultaneously.  Therefore, in a manner similar to that illustrated in Figure \ref{fig:tcph_example}, any time spread  can be attributed to fission chain smearing. However correlated gamma-gamma pairs have many sources of uncorrelated background that degrade the signal-to-noise of this signature. In addition to correlated fission gamma-rays, fissile material typically produces many more uncorrelated decay gamma-rays.  Depending on the amount of material present, these uncorrelated gamma-rays can create an overwhelming rate of accidental coincidences within a TCPH time window.  

Gamma-rays are also highly attenuated by the emitting fissile material itself, so they are primarily sensitive to the outer layers of an assembly.  It is therefore desirable to include at least one neutron in the correlated pair to gain sensitivity to a larger fraction of the total volume of material.

We have therefore determined that correlated gamma-neutron and neutron-gamma pairs offer the best sensitivity and signal to noise.  The gamma-ray provides a clean indication of the time of a fission, while the neutron is both more penetrating and a clear indication of fission, reducing the rate of accidental coincidences.

\section{Experiment}

\begin{figure}[h!] 
 \centering
	\includegraphics[width=85mm]{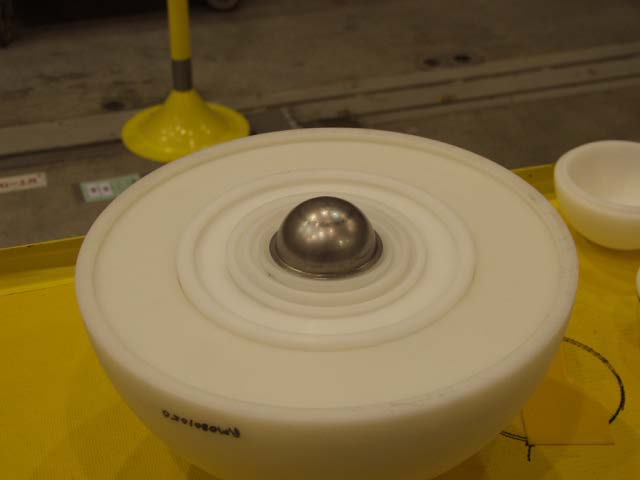}
 	\caption{Photograph of the BeRP ball nested in polyethylene hemishells, courtesy of reference \cite{Mattingly2009}.}
  \label{fig:berp_ball}
\end{figure}

The BeRP ball is a 4.5 kg sphere of $\alpha$-phase weapons-grade plutonium metal, originally manufactured in October 1980 by Los Alamos National Laboratory \cite{Mattingly2009}. This sphere has a mean radius of 3.7938 cm, and is encased in a 304 stainless steel shell that is 0.0305 cm thick. In our configurations, two identical hemishells of either polyethylene or tungsten were used that completely covered the BeRP ball. A photograph of the BeRP ball with polyethylene hemishells is shown in Figure \ref{fig:berp_ball}. 

A series of measurements of the BeRP ball were conducted at the Nevada National Security Site (NNSS) to acquire data to assess TCPH distribution analysis for highly multiplying assemblies of fissile material. To explore a range of multiplications with different levels of moderation and reflection, we measured five configurations: bare, 1.27 cm and 2.54 cm thick close fitting shells of tungsten, and 2.54 cm and 7.62 cm thick close fitting shells of high density polyethylene (HDPE). The measurement times for each configuration were 5515, 3600, 8999, 3600, and 2509 seconds, respectively. 

\begin{figure}[h!] 
 \centering
	\includegraphics[width=85mm]{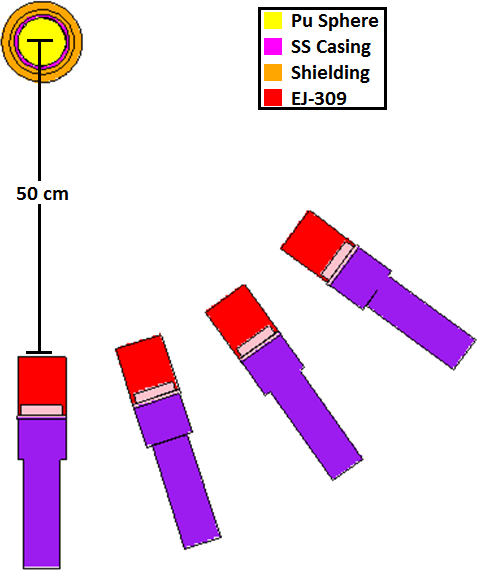}
 	\caption{Diagram of the experimental setup of the BeRP ball surrounded by four EJ-309 detectors. The detectors were spaced approximately 8.5$^{\circ}$ apart.}
  \label{fig:setup}
\end{figure}

Data were collected using four 7.62$\times$7.62  cm cylindrical EJ-309 liquid scintillation detectors. The source-to-detector distances were measured from the center of the BeRP ball. Most configurations were measured at 50 cm distances with the exception of 2.54 cm tungsten and polyethylene measured at 48 and 60 cm, respectively. A diagram of the experimental setup, taken from an MCNPX-PoliMi \cite{Pozzi2003} model, is shown in Figure \ref{fig:setup}. Anode outputs from 7.62 cm Electron Tubes photomultiplier tubes (PMTs) coupled to each cell were digitized using CAEN DT5720 digitizer, capable of 12-bit (nominal) vertical resolution and 250 MHz sampling rate. All data processing was performed off-line after the measurements, and a 0.2 MeVee (MeV electron-equivalent), equivalent to neutron energy of 1.2 MeV, threshold was applied in post-processing. 

\section{TCPH model construction}

In order to access the fission chain timing distribution, we constructed a model of TCPH distributions for multiplying sources.  Our model consists of a linear combination of the expected distribution for gamma-neutron pairs that are correlated by the same fission and those correlated by different fissions within the same chain.

The like-fission distribution is determined by a combination of the detector response to fission spectrum neutrons (Watt) and the time delay expected for the source-to-detector distance.  The different-fission distribution includes additional time smearing due to the time difference between any two fissions within a chain.  For this ``smeared" different-fission distribution the neutron has a detector response and travel time characteristic of its energy, but now a gamma-ray from a different fission starts the clock for the inter-event time.

\subsection{Like-fission TCPH distribution}

The TCPH distribution is a bivariate histogram of an estimate of neutron energy (by pulse height) and time  between correlated neutron and gamma-ray. This time is dependent on the incident neutron energy and source-to-detector distance. Therefore, for like-fission events the TCPH distribution should depend on three factors: source-to-detector distance, source neutron energy spectrum, and the energy-dependent detector response to neutrons.  Under controlled lab conditions, the source-to-detector distance can be readily measured, but it is possible to include this as a free parameter as we will later demonstrate. 

\begin{figure}[h!] 
 \centering
 \begin{subfigure}[h]{85mm} 
		\centering
		\includegraphics[width=\textwidth]{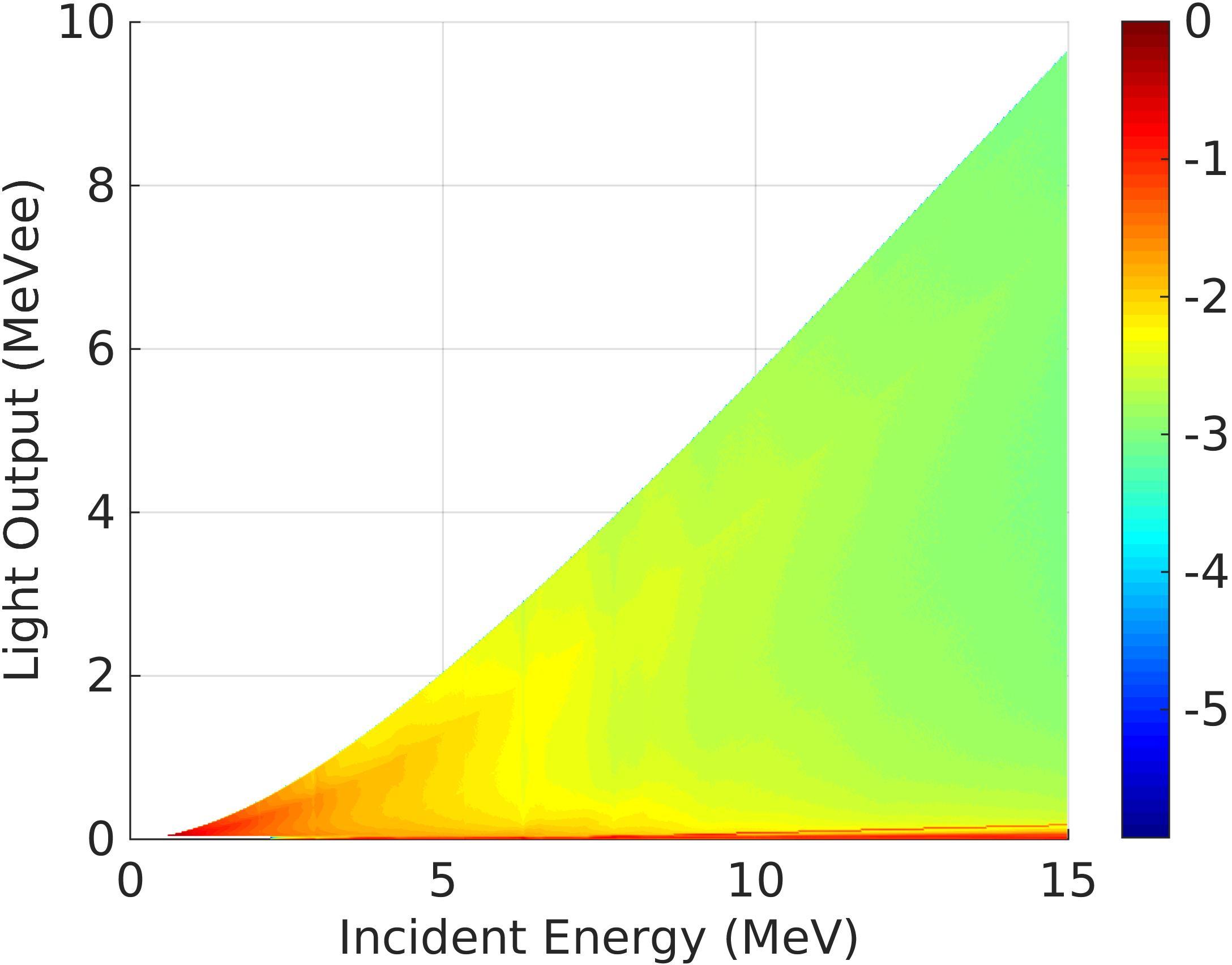}
 		\caption{Response Matrix}
  \end{subfigure}%
        ~ 
          
  \begin{subfigure}[h]{85mm} 
		\centering
	  \includegraphics[width=\textwidth]{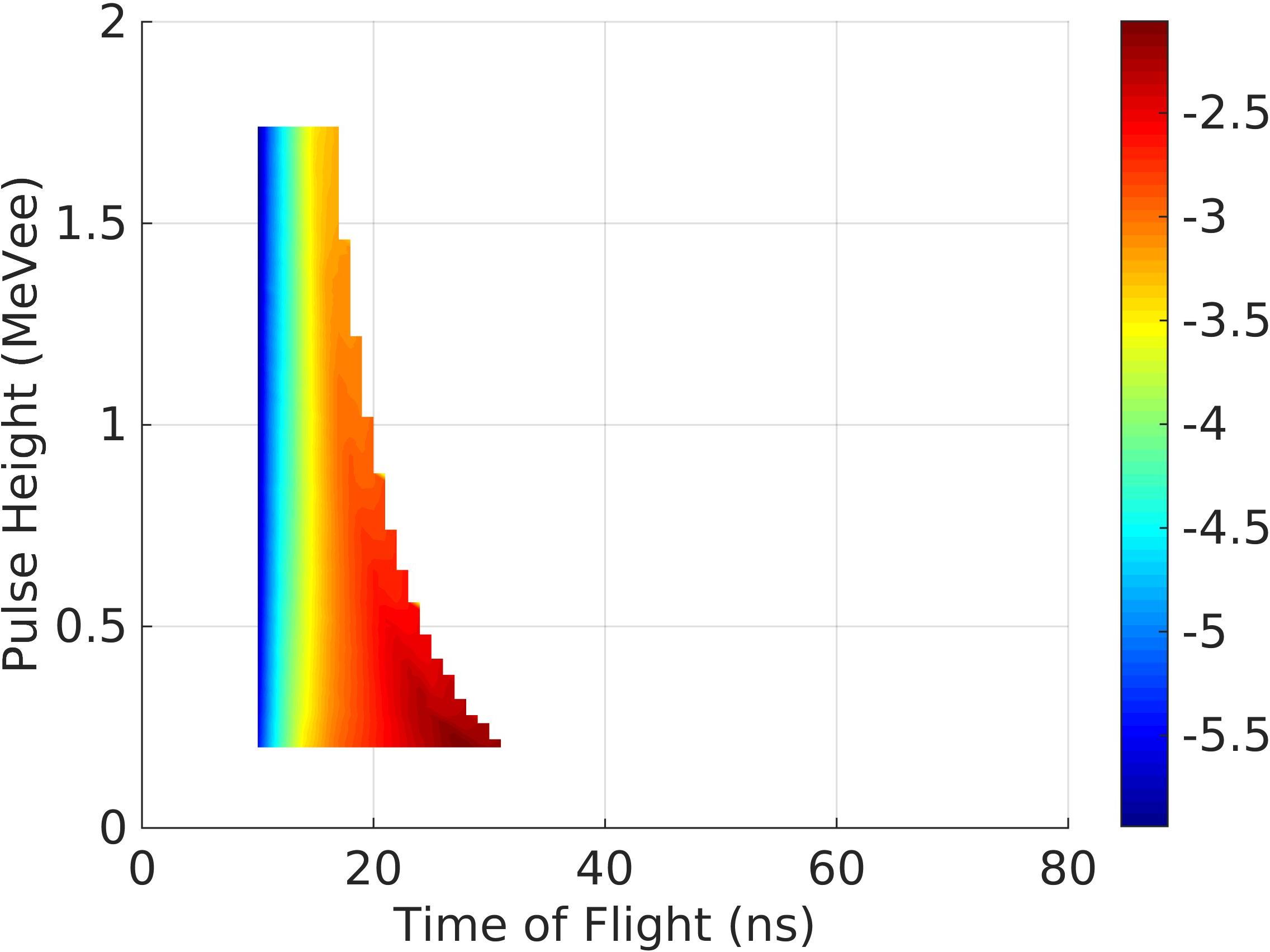}
   	\caption{TCPH Distribition}
  \end{subfigure}%
  \caption{Simulated EJ-309 neutron response matrix and the resulting TCPH distribution with assumed $^{239}$Pu Watt neutron energy spectrum and source-to-detector distance of 50 cm. Each distribution is normalized to unity and displayed on a logarithmic scale.}
  \label{fig:tcph_response}
\end{figure}

We begin with neutrons with energies sampled from the $^{239}$Pu fission Watt spectrum \cite{MCNP5}. The pulse height response of the detector is stored in a response matrix that maps an incident neutron energy to a probability density function over pulse heights. The matrix was constructed using an MCNPX-PoliMi simulation to determine the expected true energy deposition. Subsequently, a light output function was used to convert that energy into a pulse height distribution. We used a light output function of the following form:

\begin{align} \label{eq:light}
L = a E_p - b(1-\exp(c E_p))
\end{align}
where $a$, $b$ and $c$ are detector specific constants and $E_p$ is the proton recoil energy in MeV. We used parameter values for EJ-309, specified in reference \cite{Enqvist2013}, of 0.817 MeVee/MeV, 2.63 MeVee and -0.297 MeVee$^{-1}$ respectively. We found that the small variations in light output functions published in references \cite{Enqvist2013} and \cite{Lawrence2014} had no discernible impact on the resulting TCPH distributions and our fitted parameters. The simulated neutron response matrix and the resulting TCPH distribution are shown in Figure \ref{fig:tcph_response}.

\subsection{Fission chain timing distribution}

\begin{figure}[h!] 
 \centering
	\includegraphics[width=85mm]{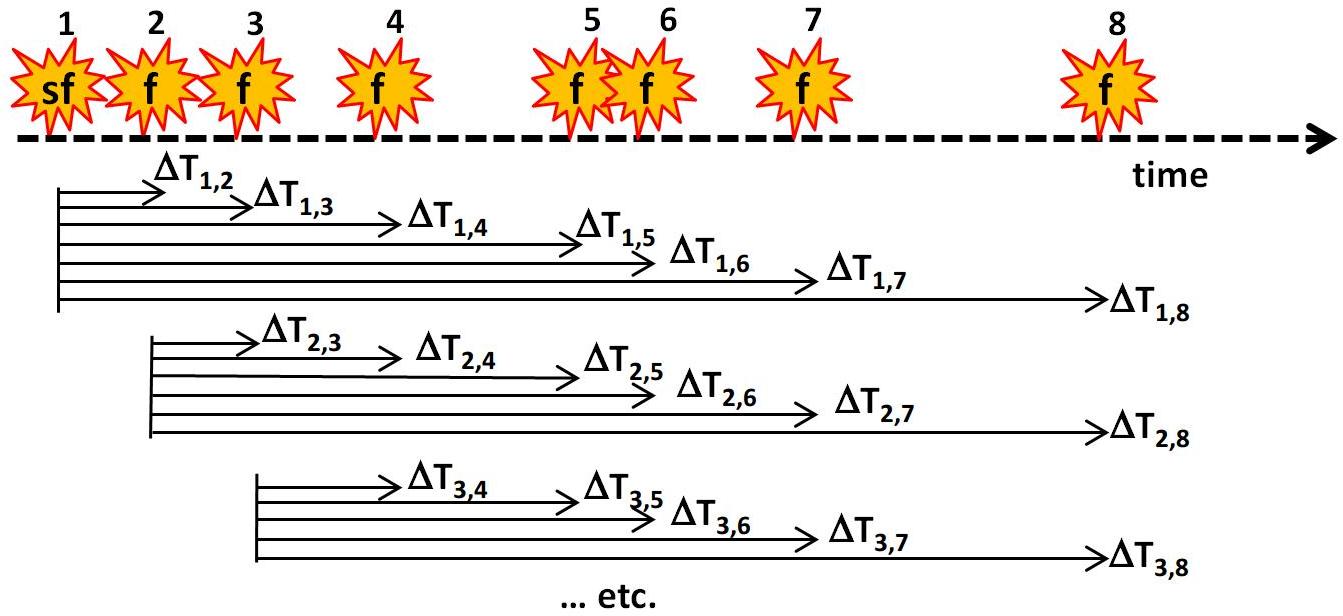}
 	\caption{Time differences between n$^{th}$-nearest-neighboring fission events in a chain starting with the spontaneous fission (sf), followed by subsequent induced fissions. The correlation of all possible event pairs from a fission chain contribute to the spread in the TCPH distribution from multiplying sources. Since it is possible to correlate a gamma-ray with a neutron from either a previous or future fission event in a chain, the TCPH distribution is smeared in both time directions.}
  \label{fig:fission_chain}
\end{figure}

In order to include the effect on the TCPH distribution caused by gamma-rays and neutrons being correlated from different fissions in the same chain, we posit that the time spread is driven by the distribution of time difference between all possible pairs of events in a fission chain, as illustrated in Figure \ref{fig:fission_chain}. This distribution will be governed by multiple physical attributes, including sub-critical multiplication which is defined as 

\begin{align} \label{eq:multiplication}
M = \frac{1}{(1-k)}
\end{align}
where $k$ is the ratio of the number of neutrons in any generation to the previous one. Multiplication directly corresponds to the average length of a fission chain. The presence of neutron moderating materials, such as polyethylene, will stretch the time between fission events as the neutrons inducing fission are moderated and reflected. 

As stated previously, a detailed physical model of the fission chain process and how it responds to these factors is beyond the scope of this paper. There has been a lack of demand to efficiently model the actual timing between fission events in a chain. For reactor analysis the principal aspect of the fission chain is its average length and time-dependent analysis (point reactor kinetics) is typically limited to the rise and fall of total neutron population in a system. As a result an off-the-shelf tool that could efficiently model fission chain timing dynamics is not readily available. Instead, we introduce a gamma distribution as an empirical substitute for a physical model.  

\begin{figure}[h!] 
 \centering
	\includegraphics[width=85mm]{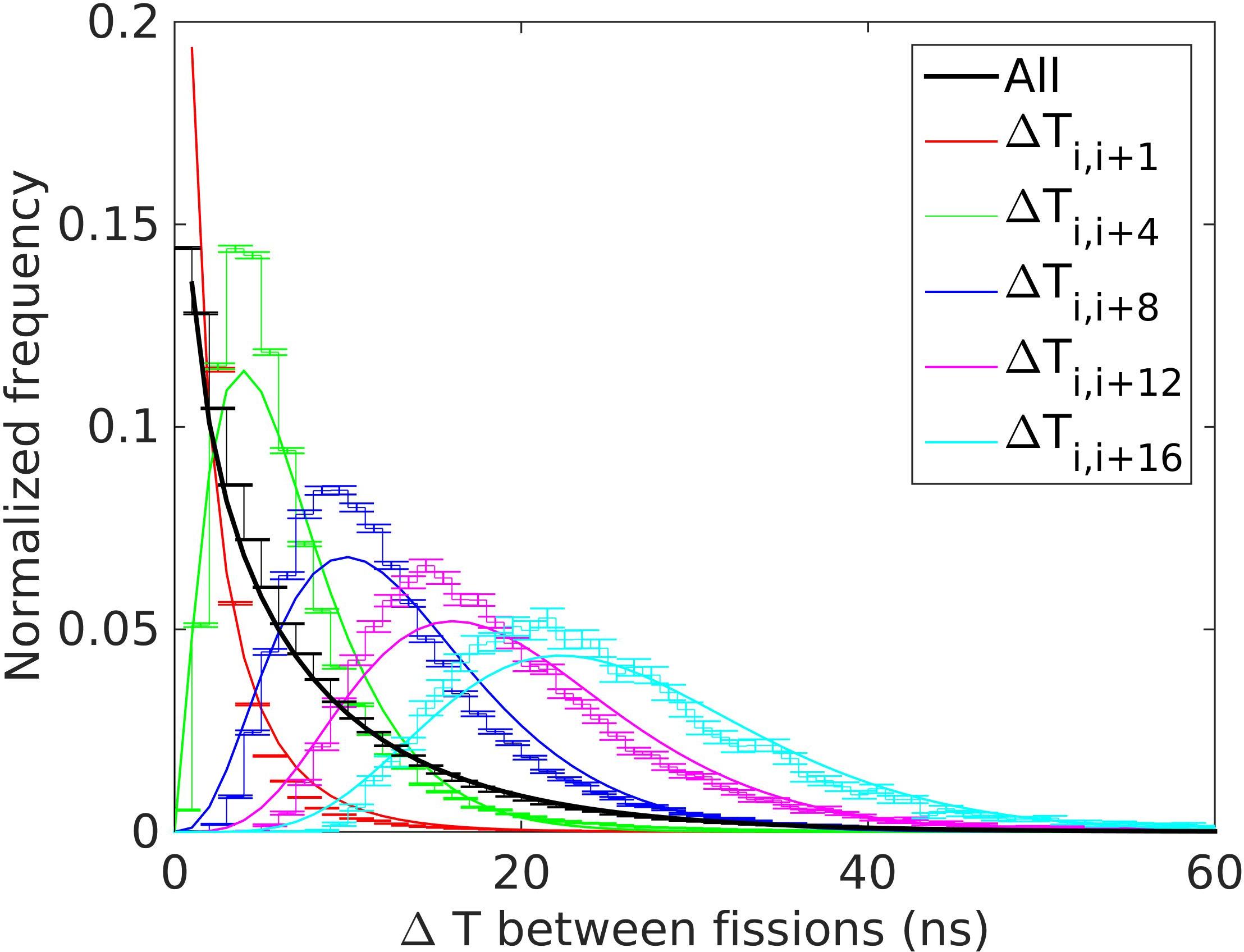}
 	\caption{Distribution of time differences for various n$^{th}$-nearest-neighbor fission events is represented by each colored line, with the black line showing the distribution between all fission events in a chain. The distributions were obtained from an MCNPX-PoliMi simulation of bare BeRP ball (stairs) and then normalized to unity and fit by gamma distributions (solid lines).}
  \label{fig:sim_fiss_chain}
\end{figure}

The probability density function of a gamma distributed random variable is

\begin{align} \label{eq:gamma}
f(x) = \frac{1}{\Gamma(\alpha) \theta^{\alpha}} x^{\alpha-1} e^{-x/\theta}
\end{align}
where $\alpha$ is the shape parameter and $\theta$ is the rate parameter. Two properties of the gamma distribution make it a plausible representation for the distribution of times between fission events. First, the gamma distribution describes the waiting times until $\alpha^{th}$ Poisson distributed event. Second, the sum of independent gamma distributions follows a gamma distribution. The independence condition is not true for events in a fission chain; however, we can empirically demonstrate the time distribution between fission events roughly follows a gamma distribution by comparing MCNPX-PoliMi simulations. The simulated distribution of time differences between fission events in a bare BeRP sphere and corresponding gamma distribution fits are shown in Figure \ref{fig:sim_fiss_chain}. Although the time distribution between exact number of fission events follow a gamma distribution, with coefficient of determination (R$^2$) between 0.930 and 0.949, the fit for all possible fission combinations has R$^2$ of 0.997. 

The results from the bare BeRP case are encouraging, but there are several important caveats to consider. Unsurprisingly, the goodness of fit of the gamma distribution deteriorates for configurations with increasing amount of moderator. There is nothing in the gamma distribution that captures the effect that reflected neutrons can have on the fission chain distribution. Most importantly, however, the simulation in Figure \ref{fig:sim_fiss_chain} includes the time between fissions throughout the fissile material, but neglects the effects of attenuation within the material on particles that escape and are available for detection. 

\subsection{TCPH distribution for multiplying sources}

The TCPH distribution for a multiplying source is constructed using a linear combination of the non-multiplying TCPH distribution, as shown in Figure \ref{fig:tcph_response}(b), and a TCPH distribution smeared by the gamma distribution.  The smearing is accomplished by setting each bin on the TCPH distribution equal to the sum of neighboring bins along the time axis, with a weight given by the gamma distribution for the time delay between bins. Included in the sum is the contribution of the original same fission bin weighted by factor $n$. 

The factor $n$ represents the fraction of correlated pairs from the same fissions over all fission and in theory should be proportional to the length of fission chain $L$:
\begin{align} \label{eq:prop_n}
  n \propto \frac{L}{ \binom{L}{2} }.
\end{align}
By definition the average length of the fission chain is related to the sub-critical multiplication from Eq. \ref{eq:multiplication} such that $M = (\bar{L} - 1)$. 

To account for $n$ in our model we add a conditional statement to the gamma distribution: 
\[
F(x|\alpha,\theta,n) = 
  \begin{cases}
    f(x|\alpha,\theta) &\quad\text{if x} > 0 \\
    n  &\quad\text{otherwise.} \
  \end{cases}
\]
where $f(x)$ is defined in Eq. \ref{eq:gamma}. The smeared TCPH distribution is calculated by 

\begin{align} \label{eq:smear}
\bm{M} = \bm{N} \times F(\bm{A}) 
\end{align}
where $\bm{N}$ is a pulse height by time matrix of the like-fission TCPH distribution. The matrix $\bm{A}$ is a symmetric Toeplitz matrix with the first row defined as the cumulative difference between the times in the TCPH distribution $\bm{N}$. In this case, the function $F$ performs a point-wise operation on values in matrix $\bm{A}$.

The resulting TCPH distribution $\bm{M}$ depends on $\alpha$, $\theta$ and $n$ which define how the distribution is smeared, and the distance from the detector which shifts the non-multiplying TCPH distribution $\bm{N}$. If the distance is known it can be fixed, otherwise it can be included as a free parameter with the other three parameters. To solve for the parameters we define the loss function as the root-mean-square error (RMSE) between the modeled TCPH distribution $\bm{M}$ and a measured distribution. We then minimize this function by employing an unconstrained non-linear optimization method based upon the Nelder-Mead simplex algorithm \cite{Nelder1965, Lagarias98}. 

We solve for the parameters of interest in three steps in order to ensure the best-fit parameters. In the first and the final third step we optimize all four parameters, and in the intermediate second step we fix the source-to-detector distance. In between the steps we change the initial guesses to the previous solutions. The Nelder-Mead simplex algorithm works best with fewer parameters, therefore by reducing their number in the middle step we hone in on the best initial guess for the gamma distribution parameters that minimize the RMSE. We decided to fix the distance parameter because it proved to be the most well constrained with the smallest relative covariance with respect to the other parameters. 

\section{Results and Discussion}

\subsection{Measurements}

\subsubsection{Best-fit parameters and their physical interpretations}

\begin{figure*}[h!] 

 \centering
  \begin{subfigure}[h!]{80mm} 
		\centering
		\includegraphics[width=\textwidth]{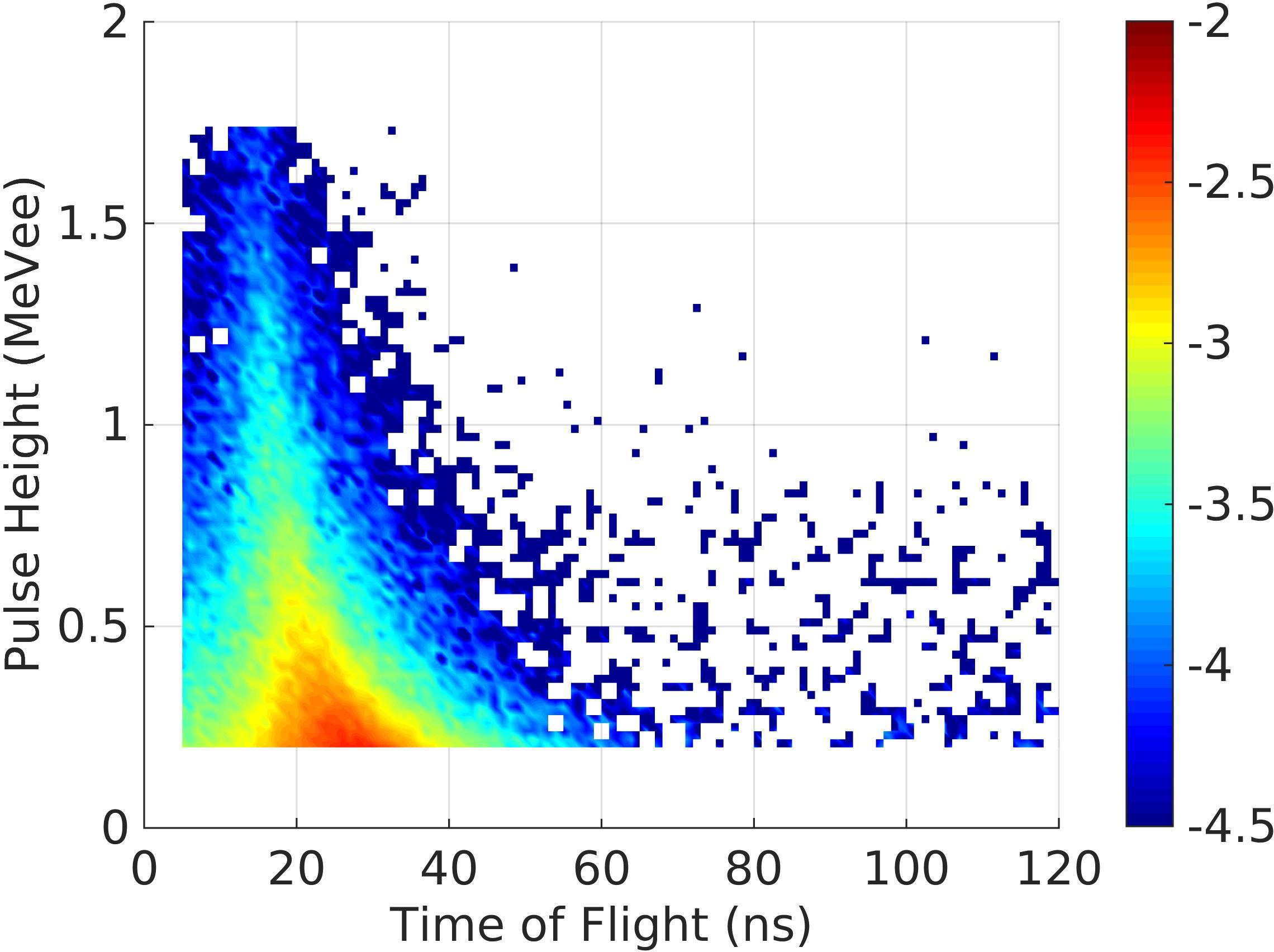}
 		\caption{Bare measured}
  \end{subfigure}%
        ~ 
  \begin{subfigure}[h!]{80mm}
		\centering
	  \includegraphics[width=\textwidth]{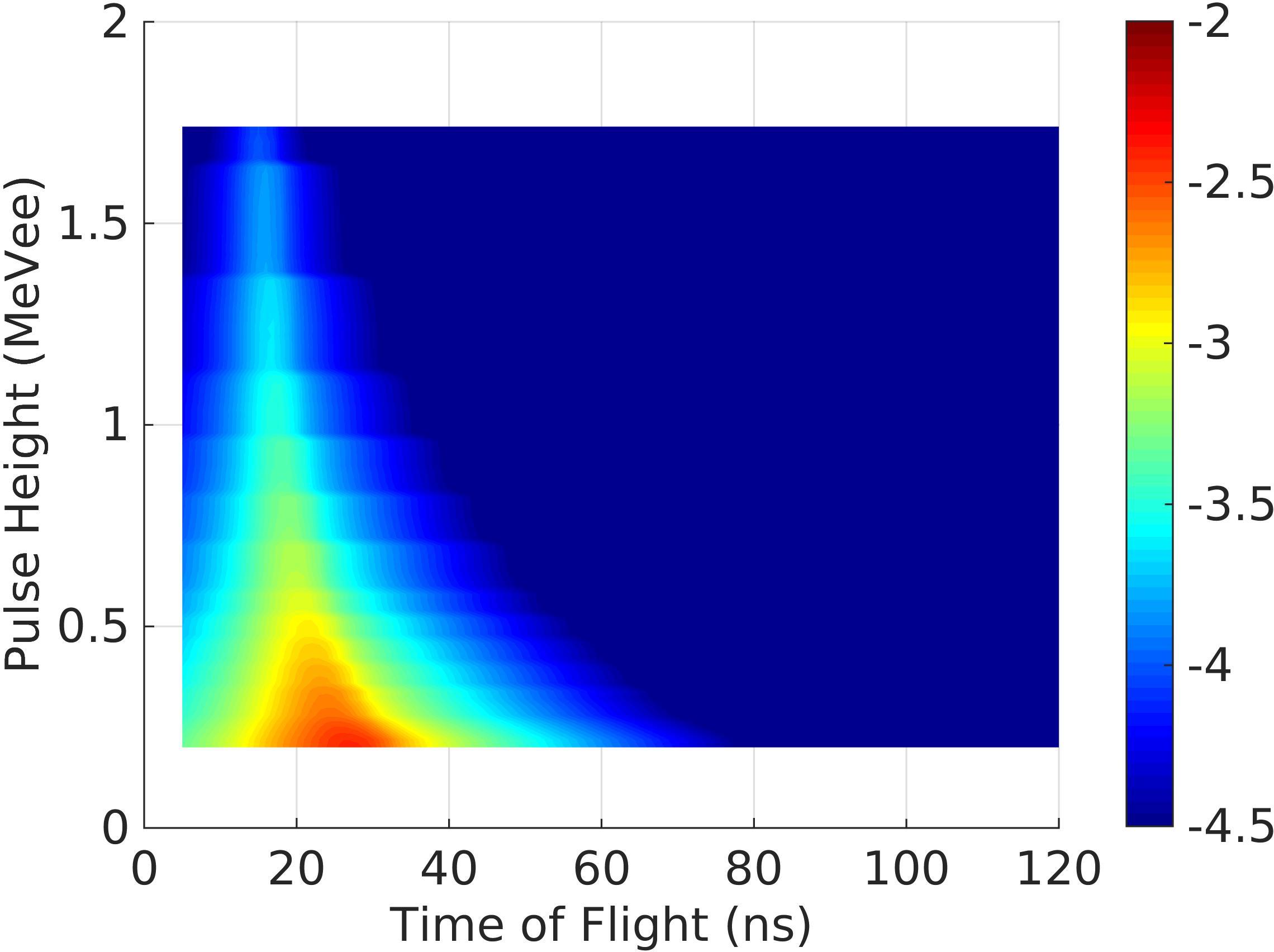}
   	\caption{Bare model}
  \end{subfigure}%
  
  \begin{subfigure}[h]{80mm} 
		\centering
		\includegraphics[width=\textwidth]{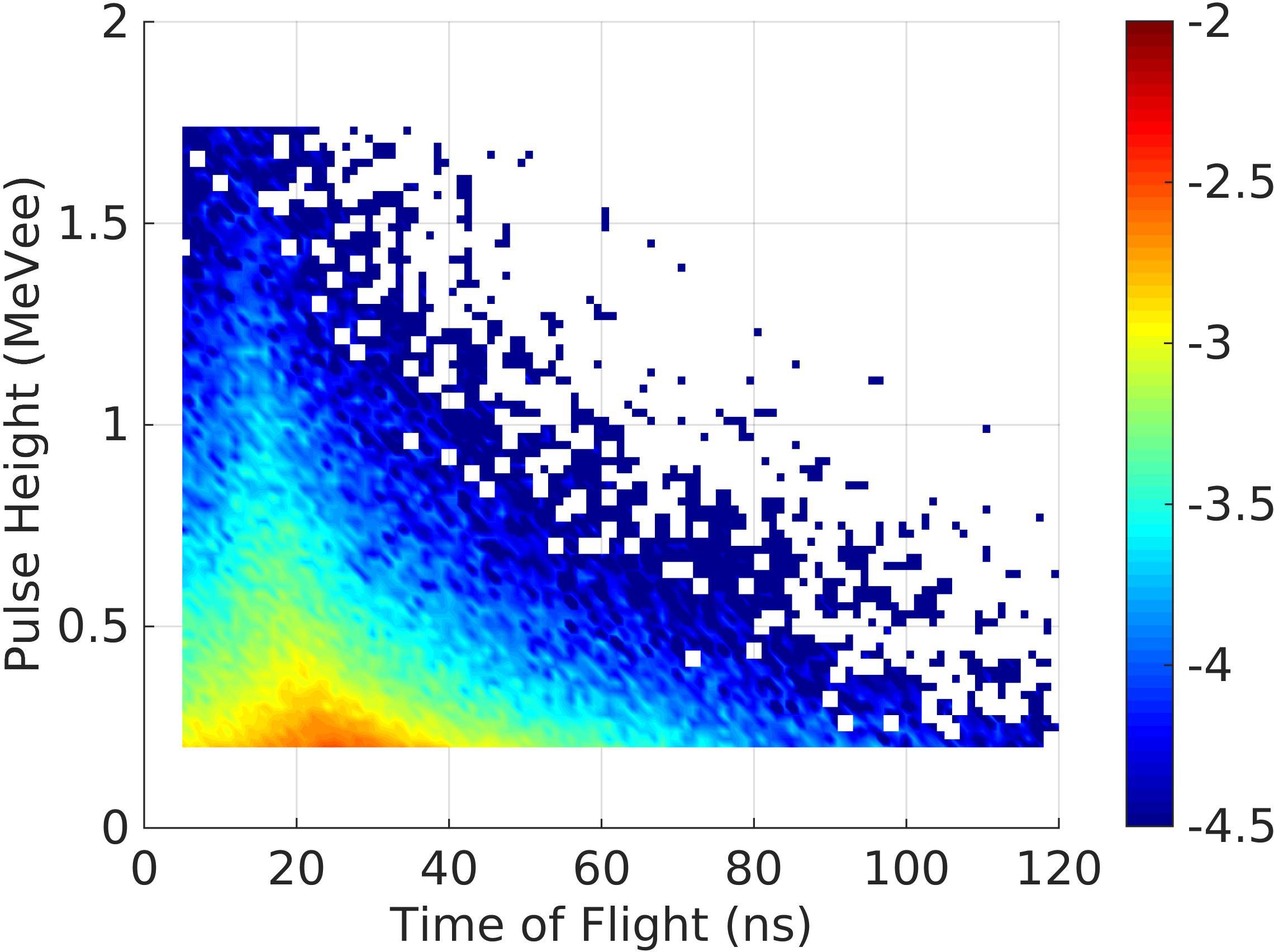}
 		\caption{2.54 cm W measured}
  \end{subfigure}%
        ~ 
  \begin{subfigure}[h]{80mm} 
		\centering
	  \includegraphics[width=\textwidth]{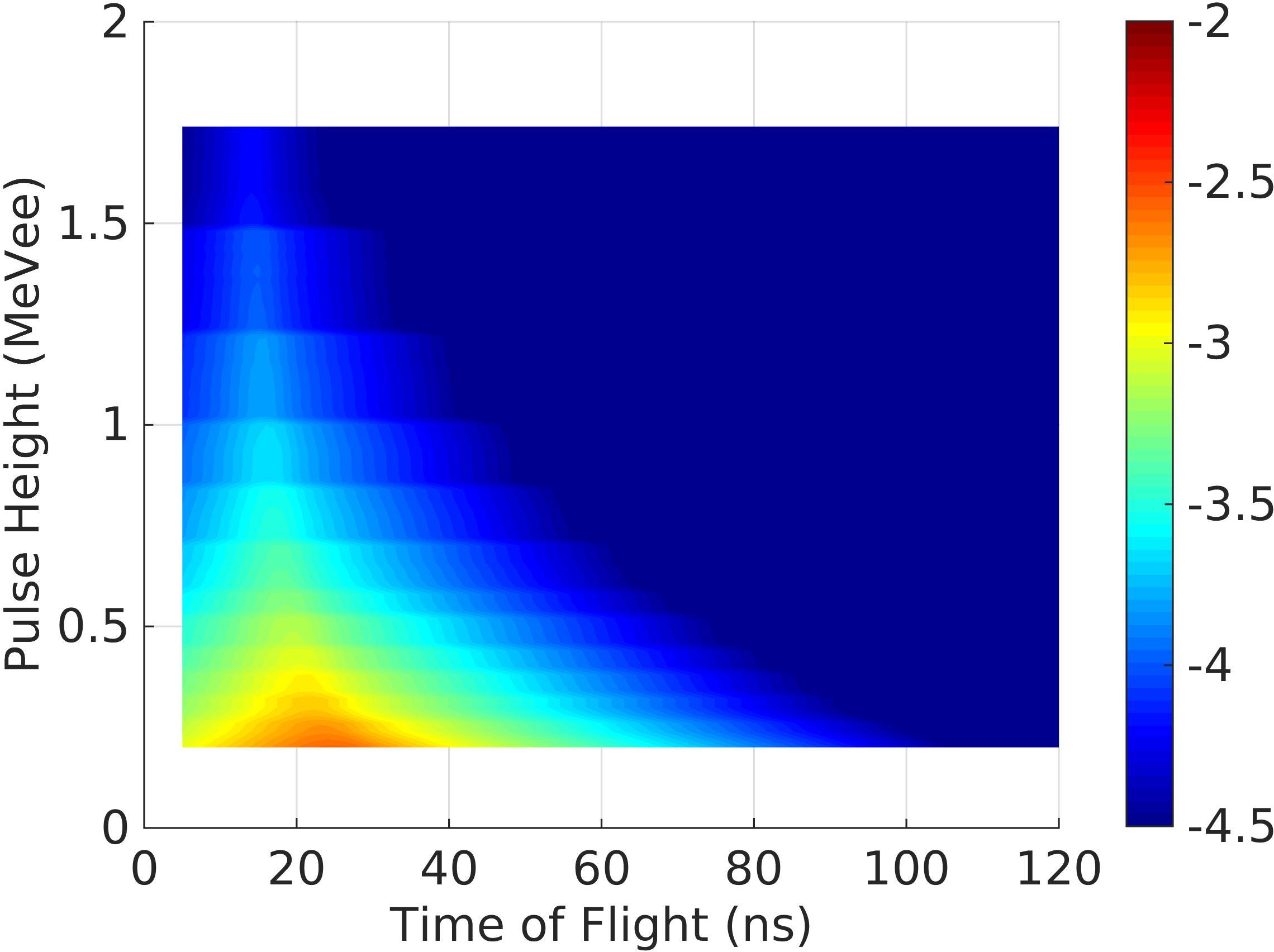}
   	\caption{2.54 cm W model}
  \end{subfigure}%
  
  \begin{subfigure}[h]{80mm} 
		\centering
		\includegraphics[width=\textwidth]{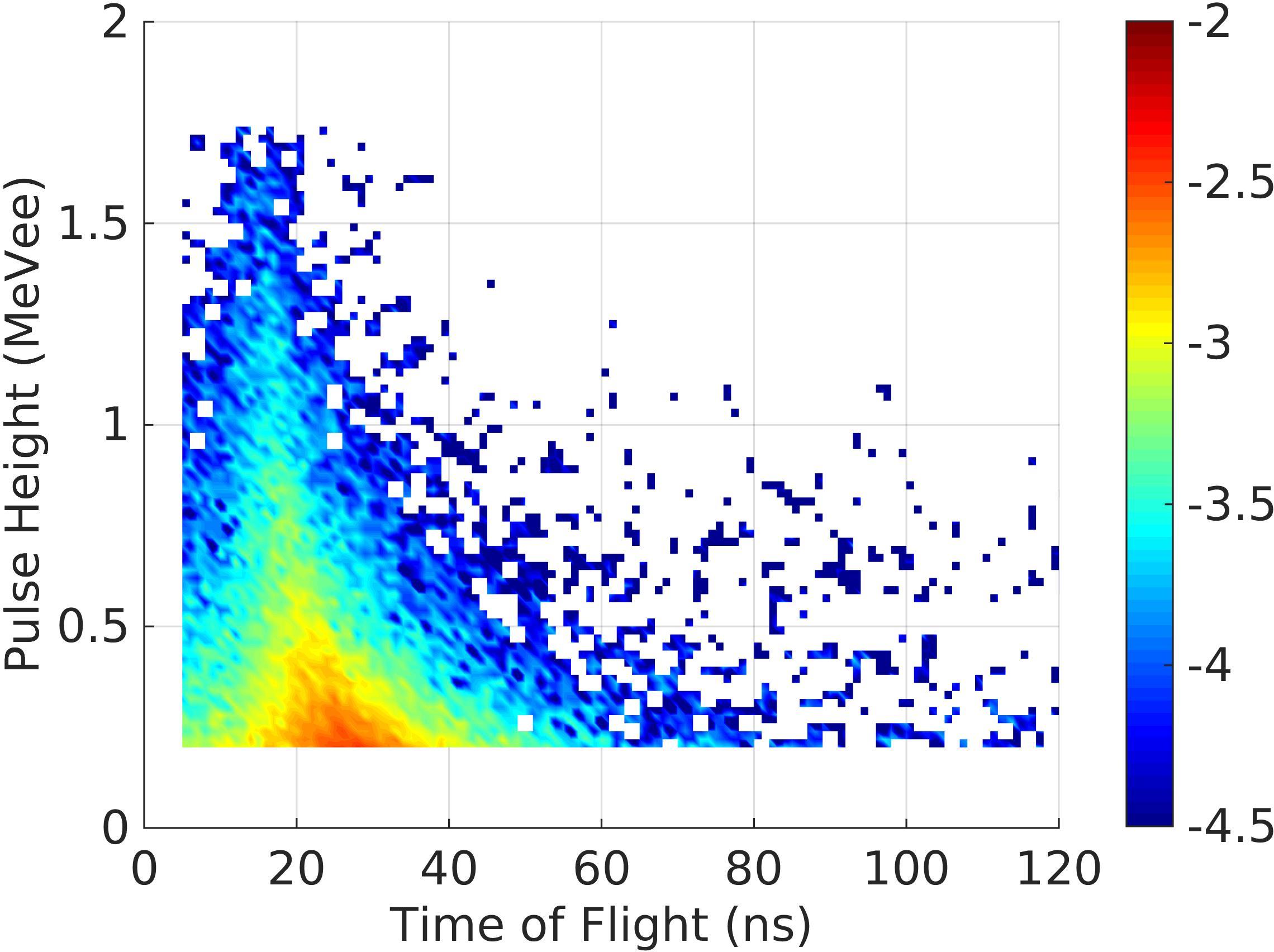}
 		\caption{7.62 cm HDPE measured}
  \end{subfigure}%
        ~ 
  \begin{subfigure}[h]{80mm} 
		\centering
	  \includegraphics[width=\textwidth]{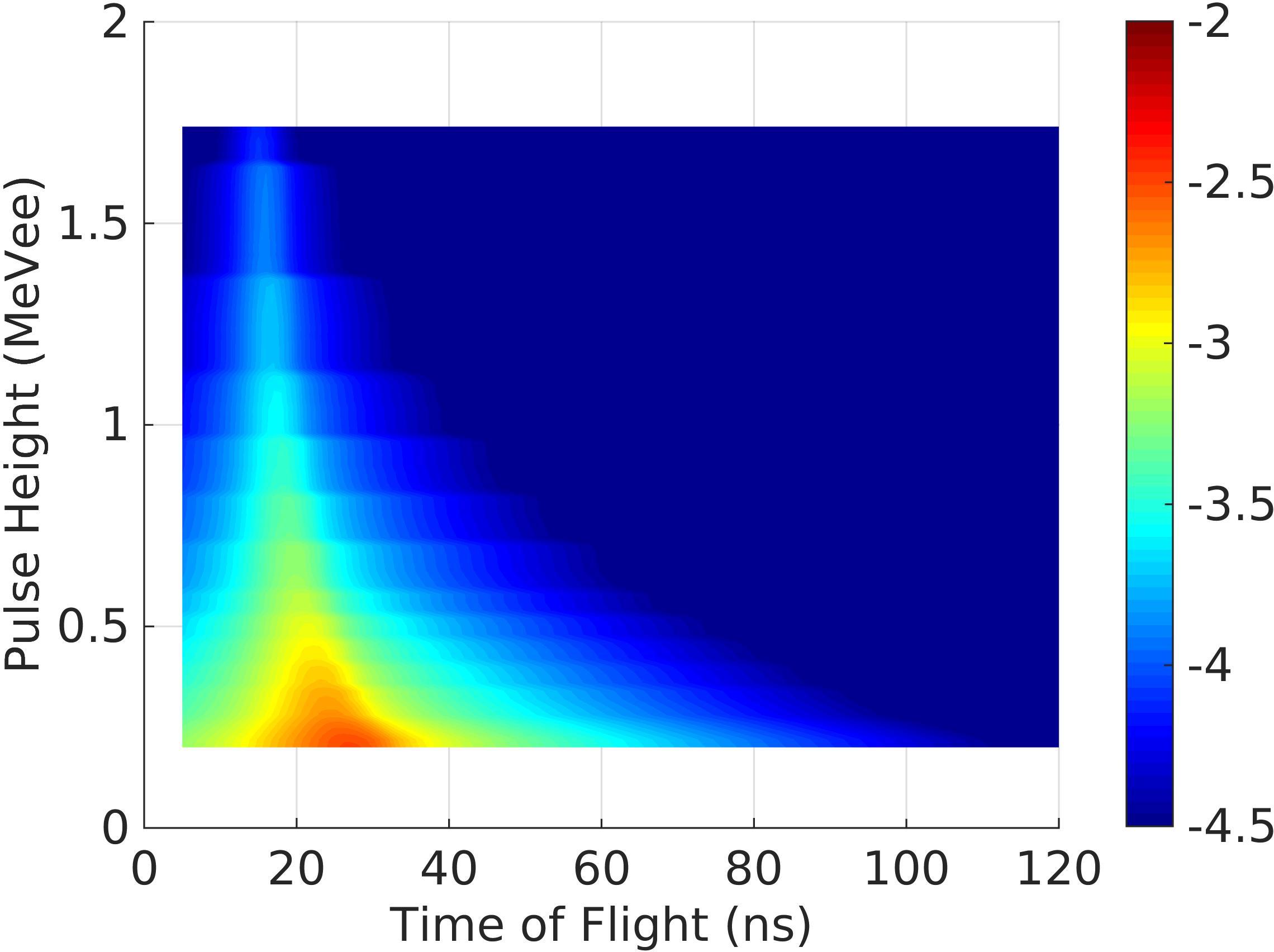}
   	\caption{7.62 cm HDPE model}
  \end{subfigure}%

  \caption{Measured TCPH distributions (left) and corresponding models with parameters from the minimization algorithm (right) for various BeRP ball configurations. Each distribution is normalized to unity and displayed on a logarithmic scale.}
  \label{fig:tcph_meas_sim}
\end{figure*}

A subset of the measured TCPH distributions and corresponding optimized models built from best-fit parameters are shown in Figure \ref{fig:tcph_meas_sim}. The like-fission TCPH distribution, shown in Figure \ref{fig:tcph_response}(b), is smeared to produce TCPH distributions shown in subfigures (b), (d) and (f) of Figure \ref{fig:tcph_meas_sim}. The smearing parameters are optimized in order to match the measured TCPH distribution shown in subfigures (a), (c) and (e) of Figure \ref{fig:tcph_meas_sim}.

The optimized gamma distribution parameters and factors of $n$ are shown in Table \ref{table:solutions_param}. We expect that the parameter $n$ would decrease with increasing sub-critical multiplication given its relationship to fission chain length in Eq. \ref{eq:prop_n}. This trend is apparent within each shielding material configuration, but is not present between them. Furthermore, the relative uncertainty in this parameter is over five times greater than the others, which limits its predictive capability. 

The shape parameter, $\alpha$, clearly demarcates the differences between the polyethylene moderator and tungsten reflector. Although $\alpha$ has no obvious physical interpretation when its a non-integer less than 1, it is valuable that it can distinguish between intervening material type independently of multiplication. 

We would expect that the rate parameter, $\theta$, would increase with multiplication because it is proportional to the degree of smearing of the TCPH distribution. The corollary to this is the measure of the gradient of the TCPH distribution found in \cite{Miller2013}. This positive correlation is only apparent for the tungsten case, and is actually negative for the polyethylene. However, in Section \ref{sec:sim} we show that for simulated polyethylene cases the rate parameter is positively correlated with multiplication, as expected. 

\begin{table*}[t]
\caption{Optimized gamma distribution parameters for the five measured configurations of the BeRP sphere with standard errors of 1 standard deviation shown.}
\centering
\begin{tabular}{l| c c c c}
\hline
\textbf{case} & \multicolumn{1}{c}{\textbf{$\alpha$}} & \multicolumn{1}{c}{\textbf{$\theta$}} & \multicolumn{1}{c}{\textbf{$n$}} & \multicolumn{1}{c}{Multiplication}   \\ \hline
bare         & 0.57 $\pm$ 0.04 & 12.41 $\pm$ 0.62 & 0.08 $\pm$ 0.03 & 4.429 $\pm$ 0.002\\ 
1.27 cm W    & 0.87 $\pm$ 0.04 & 14.91 $\pm$ 0.63 & 0.11 $\pm$ 0.03 & 6.447 $\pm$ 0.004 \\ 
2.54 cm W    & 0.90 $\pm$ 0.02 & 19.67 $\pm$ 0.53 & 0.06 $\pm$ 0.01 & 8.752 $\pm$ 0.006 \\ 
2.54 cm HDPE & 0.48 $\pm$ 0.03 & 26.74 $\pm$ 1.38 & 0.08 $\pm$ 0.03 & 7.743 $\pm$ 0.005\\
7.62 cm HDPE & 0.53 $\pm$ 0.03 & 22.54 $\pm$ 0.91 & 0.07 $\pm$ 0.02 & 20.3 $\pm$ 0.2 \\ \hline
\end{tabular}
\label{table:solutions_param}
\end{table*}

The optimized source-to detector distances are shown in Table \ref{table:solutions_dist}. Multiplication values were calculated from MCNP5 simulations of each configuration \cite{MCNP5}. In all cases the optimized distance was smaller than the measured distance, with the greatest discrepancy present in the thicker tungsten case. This suggests that this effect may be due to self-shielding; correlated particles from fission events closer to the surface of the BeRP ball on the side of the detectors are more likely to be detected. Therefore, any additional high-Z material would preferentially select for gamma-rays born at the surface of the BeRP ball, which would decrease the optimized source-to-detector distance. We see this effect in simulation, where the source-to-detector distance decreases significantly for increasing tungsten reflector thickness but remains only slightly altered for low-Z polyethylene moderator.

\begin{table*}[t]
\caption{Optimized distances for the five measured configurations of the BeRP sphere with standard errors of 1 standard deviation. The actual distance was measured from the center of the BeRP ball to the face of the detectors.}
\centering
\begin{tabular}{l| c c}
\hline
\textbf{case} &  \multicolumn{1}{c}{optimized distance (cm)} & \multicolumn{1}{c}{actual distance (cm)} \\ \hline
bare         & 49.02 $\pm$ 0.06 & 50 $\pm$ 0.5 \\ 
1.27 cm W    & 49.04 $\pm$ 0.12 & 50 $\pm$ 0.5 \\ 
2.54 cm W    & 44.75 $\pm$ 0.11 & 48 $\pm$ 0.5 \\ 
2.54 cm HDPE & 59.72 $\pm$ 0.09 & 60 $\pm$ 0.5 \\
7.62 cm HDPE & 49.25 $\pm$ 0.09 & 50 $\pm$ 0.5 \\ \hline
\end{tabular}
\label{table:solutions_dist}
\end{table*}

\clearpage

\subsubsection{Discrepancies in the polyethylene cases}

Experimental setup differences and the physics of moderating neutrons offer a couple of explanations for the unexpected lower rate parameter of the 7.62 cm polyethylene case compared with the thinner 2.54 cm polyethylene case. First, the moderator has the effect of slowing the average time between fissions to a microsecond time-scale, which makes true correlations difficult to distinguish from uncorrelated background. The effect that additional polyethylene has in increasing multiplication is therefore not captured within the 120 ns measurement correlation time window. The diminishing returns on sensitivity of the TCPH distribution to multiplication with increasing layers of polyethylene moderator has previously been demonstrated in simulation \cite{Miller2012}. 

Furthermore, the thinner polyethylene case was measured 10 cm further away, and on a different experimental campaign than the other polyethylene case. A farther source-to-detector distance has the effect of broadening the TCPH distribution, which should be captured by our model. However, it could be that there are nuisance parameters due to the differences between the two experiments that are inexplicably contributing to this discrepancy. Ideally those would be teased out with additional experiments, but with limited access to the NNSA and the BeRP ball we chose to supplement our findings with simulation results. 

\subsection{Simulations} \label{sec:sim}

\begin{figure} 
  \centering
	\includegraphics[width=85mm]{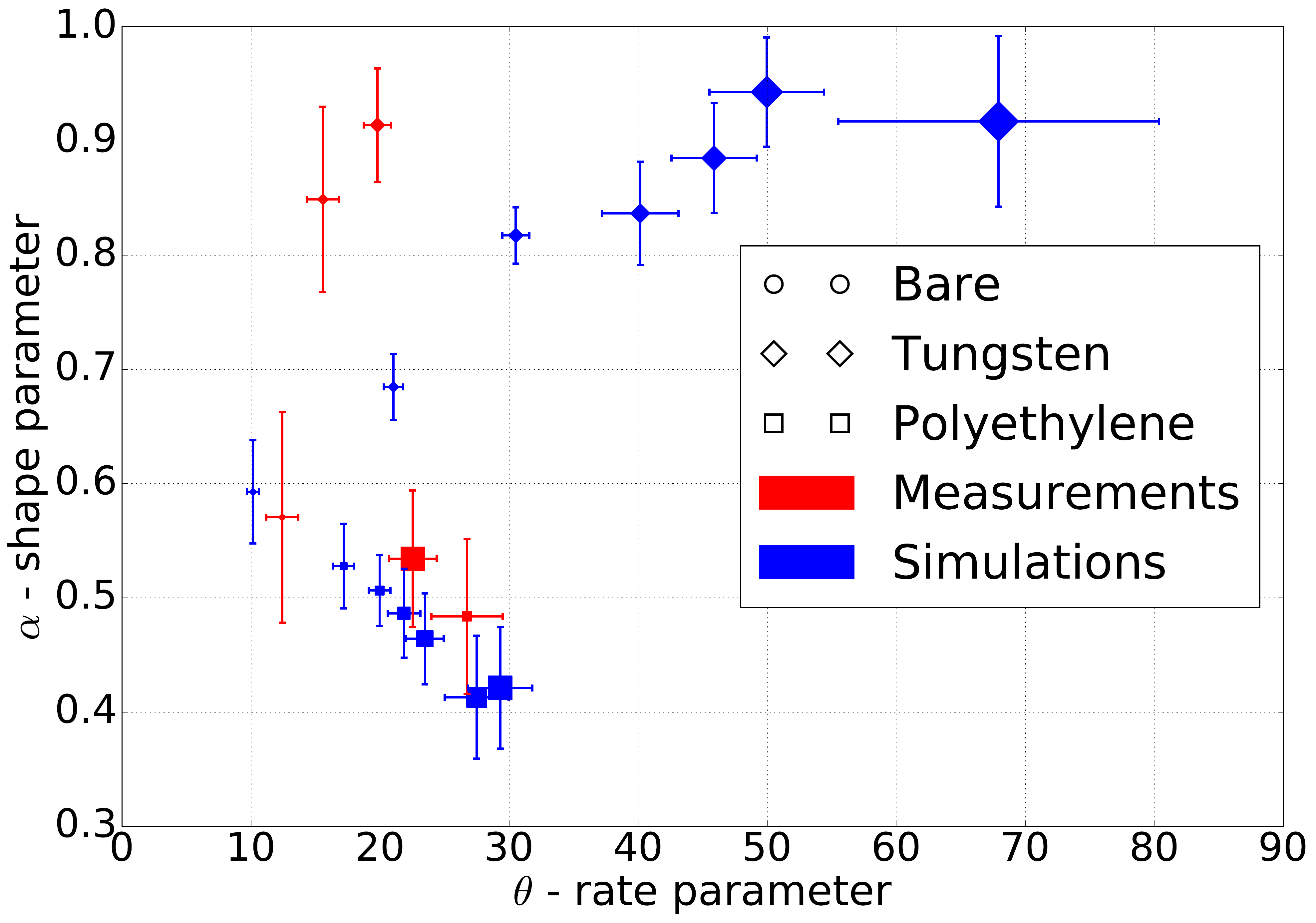}
 	\caption{Optimized shape and rate parameters for both simulation (blue) and measurement (red) cases. The shielding configurations varied in thickness from 1.27-7.62 cm with 1.27 cm intervals. The increase in symbol size corresponded to an increase in multiplication, which is proportional to the area of each marker. Note that the bare configurations are furthest to the left for both measurements and simulations.}
  \label{fig:alpha_theta}
\end{figure}

MCNPX-PoliMi simulations with shielding configurations of 1.27-7.62 cm for polyethylene and 1.27-7.62 cm for tungsten in 1.27 cm intervals were simulated to better illustrate the trends in the TCPH distribution. The optimization results of the shape and rate parameter for both simulations and measurement are shown in Figure \ref{fig:alpha_theta}, with multiplication proportional to the area of each marker. 

The standard errors were taken from a covariance matrix of the best-fit parameters derived from a numerical estimation of the Jacobian. There are considerable absolute differences between measurement and simulation, but the general trends remain the same with the notable exception of the aforementioned measured polyethylene cases. The simulations reveal that it would be possible to measure the relative change in multiplication, independently of the change of surrounding moderator for a reflector or vice versa. The main driver in the size of the relative standard errors was the number of total correlated gamma-neutron pairs. This is most apparent for the thick tungsten cases which effectively shield the vast majority of fission gamma-rays. 

\section{Conclusions}

We introduced a new approach for the characterization of SNM based on a signature of temporally correlated gamma-rays and neutrons. In our approach we fit the empirical approximation of the timing distribution between fissions in a fission chain to the measured TCPH distribution. We have shown that for the BeRP ball these empirical parameters correlate with both multiplication and the type of material (e.g. low-Z moderator or high-Z reflector) coupled to the fissile assembly. This property makes this method a candidate for treaty verification applications, where confidence in warhead dismantlement is the objective. In this case warhead dismantlement would involve the removal of high explosives, which is a form of moderating material, from the fissile material of a warhead. Our reliance on a signature that is unique to SNM makes it more difficult to spoof dismantlement of fissile material. Furthermore, the signature we used can be captured with a portable set of fast organic scintillators which could be carried by an inspector. 

Though the gamma distribution was shown to reproduce the measured TCPH distributions in bare fissile assemblies, as increasing amounts of reflector were added it became a less relevant proxy for the complete effect of the fission chain dynamics. Despite this we saw that the rate parameter of the optimized gamma distribution correlated positively with multiplication. Simulation results revealed this trend, but it was not shown to be the case for the measured polyethylene cases. This multiplication-rate parameter relationship followed a different trend-line for moderated and reflected systems; which were easily identified by the shape parameter.

For this work, an empirical model was utilized to identify general trends in the fissile assemblies. However, if this were replaced by a physical model of the timing distribution of fissions within fission chains, then it may be possible to further improve the available information present in this signature.  In future work, it may be found that potential candidate models such as multi-region point-kinetics, 1D deterministic transport solvers, or even simple fission chain Monte Carlo simulations can be used to relate this signature more directly to the physical distribution of fissile material.

\section*{Acknowledgments} 

This material is based upon work supported by the U.S. Department of Homeland Security under Grant Award Number, 2012-DN-130-NF0001. The views and conclusions contained in this document are those of the authors and should not be interpreted as representing the official policies, either expressed or implied, of the U.S. Department of Homeland Security.

Sandia National Laboratories is a multi-program laboratory managed and operated by Sandia Corporation, a wholly owned subsidiary of Lockheed Martin Corporation, for the U.S. Department of Energy's National Nuclear Security Administration under Contract DE-AC04-94AL85000. SAND Number 2017-0183 J.

This work was funded in-part by the Consortium for Verification Technology under Department of Energy National Nuclear Security Administration award number DE-NA0002534.


\bibliography{mybibfile}

\end{document}